\documentclass[12pt]{article}
\usepackage{graphicx}
\usepackage{amsmath}
\usepackage{amssymb}
\usepackage[left=1in,top=1in,right=1in,bottom=1in]{geometry}
\usepackage{natbib}
\usepackage{setspace}
\makeatletter
\renewcommand\section{\@startsection{section}{1}{\z@}{-3.25ex plus -1ex minus -.2ex}{1.5ex plus .2ex}{\normalsize\bf}}
\renewcommand\subsection{\@startsection{subsection}{2}{\z@}{-3.25ex plus -1ex minus -.2ex}{1.5ex plus .2ex}{\normalsize\bf}}
\renewcommand\subsubsection{\@startsection{subsubsection}{3}{\z@}{-3.25ex plus -1ex minus -.2ex}{1.5ex plus .2ex}{\normalsize\bf}}
\makeatother

\newtheorem{thm}{Theorem}[section]

\numberwithin{equation}{section}

\newcommand{\supp}[1]{\text{supp}(#1)}

\begin{document}
\begin{center}
\textbf{On the Status of the Geodesic Principle in Newtonian and Relativistic Physics}\footnote{Thank you to David Malament and Jeff Barrett for helpful comments on a previous version of this paper and for many stimulating conversations on this topic.  Thank you, too, to helpful audiences in Paris, Wuppertal, and London, ON, especially John Manchak, Giovanni Valente, Craig Callendar, Alexei Grinbaum, Harvey Brown, David Wallace, Chris Smeenk, Wayne Myrvold, Erik Curiel, and Ryan Samaroo.  I am particularly grateful to the philosophy of physics faculty at the University of Western Ontario for awarding the 2011 Robert K. Clifton memorial prize to a previous draft of this paper.}\\[1\baselineskip]
James Owen Weatherall\footnote{weatherj@uci.edu} \\Logic and Philosophy of Science \\
University of California, Irvine
\\[1\baselineskip]
\end{center}
\singlespacing
\begin{center}\textbf{Abstract}\end{center}
A theorem due to Bob Geroch and Pong Soo Jang [``Motion of a Body in General Relativity.'' \emph{Journal of Mathematical Physics} \textbf{16}(1), (1975)] provides a sense in which the geodesic principle has the status of a theorem in General Relativity (GR).  I have recently shown that a similar theorem holds in the context of geometrized Newtonian gravitation (Newton-Cartan theory) [Weatherall, J. O. ``The Motion of a Body in Newtonian Theories.'' \emph{Journal of Mathematical Physics} \textbf{52}(3), (2011)]. Here I compare the interpretations of these two theorems.  I argue that despite some apparent differences between the theorems, the status of the geodesic principle in geometrized Newtonian gravitation is, \emph{mutatis mutandis}, strikingly similar to the relativistic case.\\
\rule{6.5in}{1pt}\\[1\baselineskip]

\doublespacing

\section{Introduction}\label{sec-introduction}

The geodesic principle is the central principle of General Relativity (GR) that describes the inertial motion of test particles.  It states that free massive test point particles traverse timelike geodesics.  There is a long-standing view, originally due to Einstein, that the geodesic principle has a special status in GR that arises because it can be understood as a theorem, rather than a postulate, of the theory. (It turns out that capturing the geodesic principle as a theorem in GR is non-trivial, but a result due to Bob Geroch and Pong Soo Jang (\citeyear{Geroch+Jang}) succeeds in doing so.)\footnote{There are many approaches to modeling the motion of test particles in the literature, but here I will focus on just this one.  For recent surveys of other approaches to the problem of describing the motion of a body in GR, see for instance \citet{Blanchet} or \citet{Damour}.  For a classic treatments of the problem, including a review of early approaches, see \citet{Dixon} and \citet{Carmeli}.  \citet{Geroch+Jang} offer brief but insightful comments about the difficulties facing some of the most intuitively obvious ways of capturing the geodesic principle in the introductory remarks of their paper.  A particularly prominent alternative approach involves the use of generalized functions, or distributions.  It seems to me that there is a close connection between the Geroch-Jang approach and the distributional approach, although exploring this connection is beyond the scope of the current paper.}$^,\,$\footnote{The Geroch-Jang theorem requires one to assume that the test particle does not contribute to the right hand side of Einstein's equation, and thus neglects any dynamical effects due to the test particle's own energy-momentum.  A more recent result \citep{Ehlers+Geroch} strengthens the Geroch-Jang result to the case where the test particle's own energy-momentum is taken into account.  I will not address the Ehlers-Geroch theorem here.} The special status of the geodesic principle is then said to make GR distinctive as a theory of spacetime structure because it in some sense explains inertial motion.  For instance, in a well-known monograph on classical field theory, Moshe Carmeli writes,
\begin{quote}\singlespacing
...[G]eneral relativity theory is somewhat unique with respect to the problem of motion.  Because the Einstein gravitational field equations are nonlinear... the motion of the sources of the field is determined by the field equations.  ... It is worthwhile to emphasize that the situation with regard to motion in other classical fields differs from the one in general relativity. ... In Newton'as theory of gravitation... the physical laws fall naturally into two independent classes. ...  The complete \emph{independence} of the dynamical laws from the field equations is a direct consequence of the \emph{linearity} of the field equation. \citep[emphasis original]{Carmeli}\end{quote}
More recently, the view has been articulated and defended by Harvey \citet{Brown}.\footnote{See also \citet{Sus}.}  As Brown puts it,\begin{quote}\singlespacing Inertia, in GR, is just as much a consequence of the field equations as gravitational waves.  For the first time since Aristotle introduced the fundamental distinction between natural and forced motions, inertial motion is part of the dynamics.  It is no longer a miracle. \citep[pg. 163]{Brown}\end{quote}\doublespacing Brown concludes, ``GR is the first in the long line of dynamical theories... that \emph{explains} inertial motion'' \citep[pg. 141]{Brown}.

I will not engage with the details of Brown's or others' views here.  Instead, I want to consider a question raised in the quoted remarks, regarding the precise status of inertial motion in Newtonian physics.  In Newton's own formulation of his theory (what I will call ``standard Newtonian gravitation''), inertial motion is captured by Newton's first law of motion, which certainly does appear to have the status of a postulate.\footnote{For a detailed and enlightening discussion of the status of the first law of motion in standard Newtonian gravitation, see \citet{Earman+Friedman}.}  Here, however, I will focus on a reformulation of Newtonian gravitational theory originally due to \'Elie \citet{Cartan1, Cartan2} and Kurt \citet{Friedrichs} called ``geometrized Newtonian gravitation'' (sometimes, ``Newton-Cartan theory'').\footnote{For more on geometrized Newtonian gravitation, see \citet[Ch. 4]{MalamentGR} and references therein.}  In geometrized Newtonian gravitation, much as in GR, (a) the geometrical structure of spacetime depends on the distribution of matter within spacetime, and conversely (b) gravitational effects are seen to be manifestations of the resulting geometry.  Of particular interest for present purposes is that in geometrized Newtonian gravitation, inertial motion is captured by a geodesic principle.  In a recent paper, I have shown that in geometrized Newtonian gravitation, as in GR, the geodesic principle can be captured by a theorem.  Indeed, the formulation of this theorem is quite similar, at least mathematically, to the Geroch-Jang theorem \citep{WeatherallJMP}.

The present paper is a comparative study of the status of the geodesic principle in GR and in geometrized Newtonian gravitation, in light of the two theorems noted above.  I will begin by describing the Geroch-Jang theorem and its natural interpretation.  Then I will discuss the corresponding theorem in the Newtonian case, followed by a discussion of what I take to be the principal (potential) differences between the two theorems.  I will argue that the status of the geodesic principle in geometrized Newtonian gravitation is, \emph{mutatis mutandis}, strikingly similar to the relativistic case, though in neither theory is the situation as clean as Brown (or Einstein) may have liked.  One principal moral will be that, if one is going to argue that GR is special or distinctive regarding its treatment of inertia, the distinction lies with the status and interpretation of the conservation of energy-momentum (mass-momentum, in the Newtonian case), and not with the status of the geodesic principle \emph{per se}. I will conclude with some brief morals concerning the logical structure of physical theories.

\section{\citet{Geroch+Jang} and the status of the geodesic principle in GR}

The principal difficulty in precisifying claims that the geodesic principle is a theorem of GR is that standard statements of the principle make it difficult to identify a candidate statement for proof.  On its face, the principle is at conceptual odds with GR, wherein matter is represented by a smooth field on spacetime, the energy-momentum tensor $T^{ab}$.\footnote{Here and throughout the paper, I assume the reader is familiar with the mathematics of GR and geometrized Newtonian gravitation.  The notation and conventions used here follow \citet{MalamentGR}.} It is not immediately clear how to construct a massive point particle starting from such a field, or how to describe its dynamics from the field equations.  The Geroch-Jang theorem tackles the problem in a particularly clever way.  The idea is to start with a curve in spacetime, rather than with a body, and then show that if it is possible to construct an arbitrarily small body around the curve, then the curve must be a timelike geodesic. It follows that the only curves that arbitrarily small bodies can traverse are timelike geodesics.

\begin{thm}\label{GJ}\singlespacing \emph{\textbf{\citep{Geroch+Jang}}}\footnote{This particular statement of the theorem is heavily indebted to \citet[Prop. 2.5.2]{MalamentGR}.}Let $(M,g_{ab})$ be a relativistic spacetime, and suppose $M$ is oriented. Let $\gamma:I\rightarrow M$ be a smooth imbedded curve.  Suppose that given any open subset $O$ of $M$ containing $\gamma[I]$, there exists a smooth symmetric field $T^{ab}$ with the following properties.
\begin{enumerate}
\item \label{sdec} $T^{ab}$ satisfies the \emph{strengthened dominant energy condition}, i.e. given any timelike covector $\xi_a$ at any point in $M$, $T^{ab}\xi_a\xi_b\geq 0$ and either $T^{ab}=\mathbf{0}$ or $T^{ab}\xi_a$ is timelike;
\item \label{cons}$T^{ab}$ satisfies the \emph{conservation condition}, i.e. $\nabla_a T^{ab}=\mathbf{0}$;
\item \label{inside}$\supp{T^{ab}}\subset O$; and
\item \label{non-vanishing}there is at least one point in $O$ at which $T^{ab}\neq \mathbf{0}$.
\end{enumerate}
Then $\gamma$ is a timelike curve that can be reparametrized as a geodesic.
\end{thm}

The theorem states that if  $\gamma$ is the kind of curve that could fall within the support of a tensor field satisfying certain conditions, then $\gamma$ can be reparametrized as a timelike geodesic.  To get from here to the full statement of the geodesic principle, one wants to say that a tensor field satisfying these conditions adequately represents a free massive test point particle.  The question, then, is how adequately the restrictions imposed on the $T^{ab}$ field constructed in the theorem reflect reasonable constraints on the energy-momentum fields representing such bodies.

First, note that there is a limiting procedure implied by the set-up.  The requirement is that for \emph{any} open subset containing the image of the curve, including arbitrarily small neighborhoods, a $T^{ab}$ field exists satisfying the stated conditions.  In the presence of this limiting procedure, condition \ref{inside} captures the sense in which the object traversing the curve is a particle, rather than an extended body.  The idea is to consider curves that can be traversed by bodies described by energy-momentum tensors whose spatial support can be bounded by arbitrarily small neighborhoods of the curve.  This way of treating the problem avoids the difficulty of saying what a particle is supposed to be in GR, by instead considering arbitrarily small extended bodies.  Condition \ref{non-vanishing}, meanwhile, partially captures the sense in which the object is \emph{massive}, since it states that the energy-momentum tensor cannot vanish within the neighborhood of the curve.

Condition \ref{cons} captures the sense in which the massive particle is \emph{free}.  This interpretation of the condition is predicated on a background assumption, standard in GR, that the conservation condition always holds of the total energy-momentum tensor at a point.  If the total energy-momentum tensor is everywhere divergence free, then the only way that the $T^{ab}$ field associated with a particular matter field could have non-vanishing divergence at a point is if there are other matter fields present such that the divergence of the sum of their $T^{ab}$ fields vanishes.  Conversely, if a particular energy-momentum field is not interacting at a point, in the sense that its associated matter field is not exchanging energy-momentum with any other fields, then it must be divergence free at a point.  With this background assumption in place, the conservation condition is natural as both a necessary and sufficient condition for matter to be non-interacting.

Finally, the theorem does \emph{not} require that the field satisfy Einstein's equation, the dynamical expression governing the relationship between matter and the curvature of spacetime. The absence of such a condition indicates that the matter described in the theorem is \emph{test} matter, i.e., it is not a source term in Einstein's equation. I take it that these considerations support the desired interpretation that, at least in the presence of the limiting procedure described in the theorem and a background assumption concerning the conservation condition, a $T^{ab}$ field satisfying conditions \ref{cons}--\ref{non-vanishing} does represent a free massive test point particle.

But the theorem states an additional condition.  This last assumption, condition \ref{sdec}, can be interpreted as saying two things.  First, it says that the energy-momentum density associated with the particle must always be positive.  This condition, which corresponds to the first clause of condition \ref{sdec}, is often called the ``weak energy condition.''  So condition \ref{sdec} in part captures a second sense in which the particle is massive, complementing condition \ref{non-vanishing}.  Taken together, conditions \ref{sdec} and \ref{non-vanishing} imply that the particle has positive mass-energy, which is what is intended when one refers to a massive particle.

But condition \ref{sdec} is stronger than the weak energy condition alone.  It also says that the four momentum that any observer would associate with the particle must be timelike (whenever $T^{ab}\neq\mathbf{0}$).  In other words, it stipulates that the particle's energy-momentum must propagate strictly along timelike curves.  Condition \ref{sdec} effectively rules out two cases.  It excludes the energy-momentum fields associated with light, which, though it is a repository of energy-momentum, one nevertheless wants to think of as massless (and thus beyond the scope of the geodesic principle).  It also explicitly excludes what one might call tachyonic matter, that is, energy-momentum that can propagate along spacelike curves.

Recent work by David Malament shows that some form of energy condition is necessary for the theorem, in the sense that conditions \ref{cons} - \ref{non-vanishing} are not sufficient \citep[Prop. 2.5.3]{MalamentGR}.  More recently, I have shown that various weaker energy conditions, including the weak energy condition stated above, are not sufficient, even in the presence of the remaining three conditions \citep{WeatherallGRGP}.  These results are remarkable, at least in part because the geodesic principle is often given as the reason that GR rules out tachyonic matter fields.  Indeed, insofar as the geodesic principle is a \emph{postulate} of GR, the theory does preclude the possibility of massive particles traversing spacelike curves.  But in order to prove the geodesic principle as a \emph{theorem} of GR, one has to limit consideration to matter fields whose four-momenta as measured by any observer are always timelike.  So in a sense, the geodesic principle understood as a theorem of GR turns out to be weaker than the same statement, understood as a postulate of the theory.  The theorem requires an assumption that is emphatically not more basic than the geodesic principle itself.\footnote{One might be inclined to object at this point that the strengthened dominant energy condition is really capturing what we mean by ``massive particle,'' in the sense that massive particles should be expected to propagate along timelike curves.  But let me defer this discussion until the end of the next section, because the Newtonian theorem provides a useful contrast for understanding the strengthened dominant energy condition.}

Before proceeding to the Newtonian case, let me try to sum up what can be said about the status of the geodesic principle in GR.  I think the above discussion supports the claim that it \emph{can} be proved that free massive test point particles traverse timelike geodesics, given a background assumption regarding the conservation condition and an explicit assumption about timelike propagation.  I think that this result certainly reveals something deep about the nature of inertial motion, as Brown and others suggest---indeed, one might say that, given the above, GR explains inertial motion.\footnote{I do not intend to spell out in further detail the sense in which the above is an \emph{explanation}---instead, the position is that \emph{if} this should count as an explanation, then geometrized Newtonian gravitation \emph{also} provides an explanation of the same sort.  I might add that I am not opposed to the suggestion that these theorems should not be thought of as explanations at all---although it seems to me that they are of substantial foundational interest nonetheless, insofar as they show the deep interconnections between the various central principles of both GR and Newtonian gravitation.}   But the story is not as simple as one might hope, since the geodesic principle does not follow directly from other, more basic postulates of the theory.

\section{What is the situation in geometrized Newtonian gravitation?}

We can now move to consider the Newtonian case.  The strategy underlying the corresponding theorem in geometrized Newtonian gravitation is identical to the Geroch-Jang theorem.
\begin{thm}\singlespacing
\label{W}
Let $(M,t_a,h^{ab},\nabla)$ be a classical spacetime, and suppose that $M$ is oriented and simply connected.  Suppose also that $R^{ab}{}_{cd}=\mathbf{0}$.  Let $\gamma:I\rightarrow M$ be a smooth imbedded curve.  Suppose that given any open subset $O$ of $M$ containing $\gamma[I]$, there exists a smooth symmetric field $T^{ab}$ with the following properties.
\begin{enumerate}
\item\label{mass} $T^{ab}$ satisfies the mass condition, i.e. whenever $T^{ab}\neq \mathbf{0}$, $T^{ab}t_at_b>0$;
\item\label{cons2} $T^{ab}$ satisfies the conservation condition, i.e. $\nabla_a T^{ab}=\mathbf{0}$;
\item\label{inside2} $\supp{T^{ab}}\subset O$; and
\item\label{non-vanishing2} there is at least one point in $O$ at which $T^{ab}\neq \mathbf{0}$.
\end{enumerate}
Then $\gamma$ is a timelike curve that can be reparametrized as a geodesic.
\end{thm}
Theorem \ref{W} is as close an analogue to Theorem \ref{GJ} as could be hoped for.  Three of the conditions on $T^{ab}$ are identical, and conditions \ref{inside2} and \ref{non-vanishing2} admit the same interpretations without modification.  And once again, now in the presence of a background assumption regarding the conservation of total mass-momentum, condition \ref{cons2} also admits the same interpretation as in the Geroch-Jang theorem.

But there are also a handful of immediate differences.  First, Theorem \ref{W} requires an additional topological property---we suppose the manifold is simply connected.  The role this assumption plays is benign, however, and the theorem can be reformulated locally without this additional constraint.\footnote{There remains a subtle matter, here, that concerns the differences between the geometrical structures of the two theories, vis \`a vis integration.  However, the details are highly technical and do not bear on the current discussion.  They are described, albeit briefly, at the end of \citet{WeatherallJMP}.}  Second, Theorem \ref{W} requires a curvature condition that has no corollary in the relativistic case.  We demand that the classical spacetime in question satisfies $R^{ab}{}_{cd}=\mathbf{0}$.  This condition is necessary if one wants to recover standard Newtonian gravitation from the geometrized version of the theory. If we are interested in the geometrized version of what we antecedently thought of as Newtonian gravitational theory, this condition is not only benign, but essential for capturing the right geometrical structure.  It is only insofar as this condition holds of a classical spacetime that the spacetime is Newtonian.\footnote{For another perspective on the role of this assumption, see \citet{Sus}.  However, I think his analysis of this assumption is flawed.  I might also repeat a remark I have made previously \citep{WeatherallJMP} that it seems to me highly doubtful that this condition is in fact necessary.}

The third immediate difference concerns condition \ref{mass}.  In the Newtonian theorem, condition \ref{mass} is just the assumption that the mass density of a test object is always positive.  It is most naturally compared with the weak energy condition, and, as with the weak energy condition in the relativistic context, the mass condition along with condition \ref{non-vanishing2} should be understood to capture the sense in which the matter under consideration is  ``massive''.  But as we have seen, the strengthened dominant energy condition has an additional component, with more content than either the mass condition or the weak energy condition.

The contrast between this second part of the strengthened dominant energy condition on the one hand and the weak energy and mass conditions on the other is significant.  One might have argued that the strengthened dominant energy condition is neither surprising nor particularly strong, and that instead, it should be understood as a natural causality constraint.  The idea would be that what we mean by ``massive particle'' is an object with positive mass-energy that propagates along a timelike curve.  Fair enough.  But then the Newtonian theorem seems to be all the more remarkable, since there the geometrical structure of a classical spacetime yields timelike propagation essentially for free.  One might have thought that if any theory was going to provide a strong connection between spacetime structure and causal propagation of matter, it would be GR; and yet, at least with regard to inertial motion, it would seem that this piece of received wisdom is turned on its head.  The Newtonian theorem clearly requires less than the Geroch-Jang theorem.  And so, given this discussion, it seems right to say that, granted the background assumptions regarding the conservation of total energy-momentum and total mass-momentum, if GR can be said to explain inertial motion, then geometrized Newtonian gravitation does so at least as well.

\section{The status of the conservation condition}

Thus far, I have argued that if GR explains inertial motion, then geometrized Newtonian gravitation does as well.  But this claim relies on what I have described as a background assumption regarding the conservation condition.  Setting the difference between the energy conditions aside, the remaining point of potential difference between the two theorems concerns the status of this background assumption in each theory.  In particular, one might argue that the relevant assumption is more natural in GR than in geometrized Newtonian gravitation.

The argument I have in mind goes something like as follows.  There is a sense in which, in the general case of matter fields in GR, the conservation condition can be understood as a consequence of Einstein's field equation, $R^{ab}-\frac{1}{2} Rg^{ab}=8\pi T^{ab}$.  It is a brute geometrical fact that the left hand side of Einstein's equation is divergence free.  Hence the energy-momentum field representing the total source matter present at a point must also be divergence free.  It follows that if a source matter field is non-interacting, i.e., if it is not exchanging energy-momentum with another field at a point, then its energy-momentum field must be divergence free.

This situation is to be contrasted with the Newtonian case (the argument continues). Suppose that one begins with the classical analogue of Einstein's equation, the geometrized form of Poisson's equation, $R_{ab}=4\pi\rho t_a t_b$.  Unlike Einstein's equation, there is no way to write Poisson's equation so that one can contract one of its indices with the derivative operator.  In fact, one cannot even find a candidate expression to \emph{try} to contract with the derivative operator because there is no way to express Poisson's equation with raised indices.  Both sides of the equation simply vanish if one attempts to raise the indices with the spatial metric $h^{ab}$. And so, the conservation condition is not a constraint coming from Poisson's equation.  The upshot of the argument is supposed to be that we get the background assumption necessary to motivate condition \ref{cons} automatically in the relativistic case, but that it is an additional, independent assumption in the Newtonian case.

I take it that this kind of argument is particularly appealing to someone like Brown, who in the quoted remarks above seems to want to distinguish between the classical and relativistic cases by appealing to the explanatory role that the field equation plays in GR.  But I want to resist this argument for two reasons.\footnote{A third point of resistance might be to note that several people, notably \citet{Duval+Kunzle} and \citet{Christian}, suggest that there \emph{is} a sense in which the conservation condition can be derived from other principles in geometrized Newtonian gravitation.  It is clear that this derivation is not as straightforward as in the relativistic case, and moreover, that its foundational importance in largely unexplored.  It seems to me that a sustained discussion of the meaning and status of the conservation condition in geometrized Newtonian gravitation would be of some interest.  But it would be a digression in the present paper, and so I defer it to future work.}  The first reason is that it does not go through as simply as it seems to.  It is certainly true that Einstein's equation implies the conservation of total source energy-momentum, and Poisson's equation does not.  But insofar as we are interested in the behavior of test matter, it is not clear that this observation is to the point.  In order to get the required interpretation of condition \ref{cons} for test matter, we need an assumption that the energy-momentum fields associated with test matter are divergence free just in case the fields are non-interacting.  But we do not get \emph{that} assumption directly from Einstein's equation.  So there is a sense in which condition \ref{cons} is a bare assumption about test matter, even in the relativistic case.

One might try to save the above argument by limiting attention to a particular kind of test matter: one might argue that test matter should be understood as matter that \emph{could} be a source in the relevant equations of motion, but whose contribution to the field equations would be sufficiently small that it can be neglected.  In this case, one makes a reasonable approximation in determining spacetime curvature (say) without taking the test matter's contribution into account, yet the test matter, as a possible source field, is still constrained by the field equations.

I think this modified argument works to the extent that it gives some connection between Einstein's equation and test matter.  But if the point was to argue for the relative economy of the assumptions needed to prove the geodesic principle in GR relative to geometrized Newtonian gravitation, then I do not see how it helps much.  In the geometrized Newtonian case, one has to take the total conservation of test mass-momentum as a brute assumption.  In the relativistic case, one can either take the total conservation of test energy-momentum as an assumption, or one can assume Einstein's equation and then assume that test matter must consist exclusively of possible but neglected source matter---a new assumption that does not come up in the geometrized Newtonian case.

The second reason I want to resist the above argument is that I think there is a better way of thinking about the conservation condition.  The conservation condition is a standard assumption in both GR and in geometrized Newtonian gravitation, as well as in standard Newtonian gravitation (indeed, in four-dimensional, non-geometrized Newtonian gravitation, a simple Stokes theorem argument shows that $\nabla_a T^{ab}=\mathbf{0}$ is equivalent to ordinary conservation of momentum).  But it is also a standard assumption in a wide variety of other metric theories of gravitation, such as Brans-Dicke theory.  For this reason, I think it is most naturally thought of as a meta-principle that might be used to constrain the search for realistic  theories of gravitation.  Indeed, Einstein began with the conservation condition, and sought a field equation that was consistent with it, rather than independently discovering Einstein's equation and then deriving the conservation condition as a consequence.\footnote{See \citet{Earman+Glymour1, Earman+Glymour2} and \citet{Sauer}.  It seems that Einstein did not think of the conservation condition as something a realistic field equation would need to imply, but he did reject at least one candidate field equation because it struck him as incompatible with the conservation condition, and moreover, he considered the conservation condition as a substantive constraint on energy-momentum fields even before he recognized that the final field equation in fact implied that source matter was conserved.}  Moreover, \citet{Ehlers+Geroch}, in a follow-up paper to \citet{Geroch+Jang}, note that a virtue of theorems like the Geroch-Jang theorem is that they assume only the conservation condition, and not Einstein's equation, because this makes the results more general.

There will undoubtedly be readers for whom this alternative proposal for how to understand the conservation condition is unappealing.  But one thing seems clear, however one is inclined to think about the conservation of energy-momentum/mass-momentum.  Energy conditions aside, if one is going to maintain that GR is distinctive with regard to inertia or inertial motion, the distinction has to lie with the status of the conservation condition, and not the status of the geodesic principle, since in both cases, the geodesic principle follows from the conservation condition (among other things). Of course, if the status of the conservation condition in GR really is different than in Newtonian gravitation, then it would arguably follow that the geodesic principle also has a different status.  But it is the status of the conservation condition that is doing the work.  Indeed, one might take the principal foundational upshot of these two theorems to be that they clarify the deep connection between conservation and geodesic motion.  After all, what is ultimately shown in both cases is that (in the presence of certain energy conditions) matter will propagate along the geodesics of the derivative operator relative to which the matter is conserved.  It is the conservation of energy-momentum/mass-momentum relative to a particular derivative operator that provides the connection between the behavior of matter and the geometry of spacetime.

\section{Conclusion}

Taken as a whole, I think the comparison is a wash.  The right conclusion is that if \emph{either} theory can be thought to explain inertial motion, then \emph{both} do, in much the same way.  But more importantly, in both cases the situation is less straightforward than one might have hoped.  On the one hand, a case might be made that the conservation condition has a special status in GR.  But then one has to bend over backwards to make this special status extend to test matter, muting the point.  On the other side of the scales, to get the theorem in the relativistic case, one has to swallow a strong energy condition, with no corollary in the Newtonian case.

So what is the status of the geodesic principle in each theory?  Given how messy the details of the above discussion turn out to be, I think that to give a strong answer to this question would be a mistake.  As I have argued, there are senses in which, in both theories, the geodesic principle can be understood as a theorem.  But I think there are equally well contexts in which it is more appropriate to think of the geodesic principle as a postulate (for instance, when considering the role of tachyonic matter in GR).  Indeed, I think the idea of picking a subset of the basic principles of a theory as ``top tier'' principles, or axioms, and demoting others, is the wrong way to understand physical theories generally.  A better way of taking the results described above is to recognize that physical theories, or at least these physical theories, are founded on an interconnected network of mutually dependent principles.  It is often the case that, given a subset of these, one can prove some of the others.  But making too much of such results incurs the unacceptable cost of obscuring the often complex relations obtaining between the various parts of the theories.

\singlespacing

\end{document}